# Analysis of factors that affect students' academic performance - Data Mining Approach

Nahid Hajizadeh, Marzieh Ahmadzadeh
Department of computer engineering and IT
Shiraz University of technology
Shiraz, Iran

*Abstract—* the analysis of students' feedback can reveal imperfections and shortcomings of educational environments. The Common methods of analysis and data evaluation can't singly uncover valuable information that is hidden behind the students' feedback. This paper use two applicable data mining techniques that known as "Classification" and "Association rule discovery" to explore effective factors on non re taking a course by students.

## I. INTRODUCTION

For many students, the university study may be accompanied by depression, anxiety, severe stress, feeling mistreated by the teaching and administrative staff of the University and so on. These are the reasons for poor academic performance in students, and consequently the courses taking more than once for students will be followed. In this regard, the use of evaluation tools can be a step toward identifying and extracting factors affecting educational failure in students. Recognizing the main factors of educational failure and taking preventative and interventional strategies can help in reducing educational failure and thus improve the physical, mental and education situation of students [1].

Evaluation has long been involved in education, especially higher education domains and as one of the university management functions plays an important role in correct planning, successful implementation of educational programs and academic quality [2]. One method of evaluating effective teaching in higher education is student evaluation of courses and also faculty member's performance at the end of each semester. Students' feedback will provide specific reforms for instructor to improve the effectiveness of their performance. Mass data obtained from the evaluation, itself, would be worthless and do not come to any conclusion unless with special techniques of data mining be analyzed. Data mining techniques used in this paper are "Classification" (REPTree) and "Association rule discovery" (Apriori algorithm).

## II. EARLIER WORK

The only article that has used the same dataset [3] generally has discovered hidden patterns in data feedback from students and "Clustering" Data mining technique was applied. That paper have not performed any preprocessing on raw data and all data were analyzed together (By applying the K-means algorithm with k = 3 to 28 times per 28-attributes) and finally, the authors conclude that attributes that make similar clusters are similar in performance. This article particularly explores effective factors on non re taking a course by students, "Classification" and "Association rule discovery" techniques are applied. In this paper, all data are pre-processed before analyzing and data are grouped according to three different types of evaluations and have not been studied together. In general it can be concluded that in this paper with different view is looked at the data and although this problem initially seems to be an unsupervised problem (as in the paper [3] referred to) cleverly became a supervised problem and It seems that in comparing with the earlier paper ([3]) the result is purposeful and more accurate.

## III. DATA SET

This data set contains a total 5820 evaluation scores and 33 attributes that provided by students from Gazi University in Ankara (Turkey) [5]. Attributes nb.repeat ،attendance ،difficulty, Q1 to G12 are related to course is taken by students and attributes Q13 to Q28 are related to features of the course instructor. Specifications of all attributes are represented in table 1.

TABLE 1 SPECIFICATIONS OF ATTRIBUTES

| Name of attribute | Comment | Possible values |
|---|---|---|
| instr | Instructor's identifier | {1,2,3} |
| class | Course code (descriptor) | {1-13} |
| nb.repeat | Number of times the student is taking this course | {0, 1, 2, 3...} |
| attendance | Code of the level of attendance | {0, 1, 2, 3, 4} |
| difficulty | Level of difficulty of the course as perceived by the student | {1,2,3,4,5} |
| Q1 | The semester course content, teaching method and evaluation system were provided at the start. | {1,2,3,4,5} |
| Q2 | The course aims and objectives were clearly stated at the beginning of the period. | {1,2,3,4,5} |
| Q3 | The course was worth the amount of credit assigned to it. | {1,2,3,4,5} |

| Name of attribute | Comment | Possible values |
|---|---|---|
| Q4 | The course was taught according to the syllabus announced on the first day of class. | {1,2,3,4,5} |
| Q5 | The class discussions, homework assignments, applications and studies were satisfactory. | {1,2,3,4,5} |
| Q6 | The textbook and other courses resources were sufficient and up to date. | {1,2,3,4,5} |
| Q7 | The course allowed field work, applications, laboratory, discussion and other studies. | {1,2,3,4,5} |
| Q8 | The quizzes, assignments, projects and exams contributed to helping the learning. | {1,2,3,4,5} |
| Q9 | I greatly enjoyed the class and was eager to actively participate during the lectures. | {1,2,3,4,5} |
| Q10 | My initial expectations about the course were met at the end of the period or year. | {1,2,3,4,5} |
| Q11 | The course was relevant and beneficial to my professional development. | {1,2,3,4,5} |
| Q12 | The course helped me look at life and the world with a new perspective. | {1,2,3,4,5} |
| Q13 | The Instructor's knowledge was relevant and up to date. | {1,2,3,4,5} |
| Q14 | The Instructor came prepared for classes. | {1,2,3,4,5} |
| Q15 | The Instructor taught in accordance with the announced lesson plan. | {1,2,3,4,5} |
| Q16 | The Instructor was committed to the course and was understandable. | {1,2,3,4,5} |
| Q17 | The Instructor arrived on time for classes. | {1,2,3,4,5} |
| Q18 | The Instructor has a smooth and easy to follow delivery/speech. | {1,2,3,4,5} |
| Q19 | The Instructor made effective use of class hours. | {1,2,3,4,5} |
| Q20 | The Instructor explained the course and was eager to be helpful to students. | {1,2,3,4,5} |
| Q21 | The Instructor demonstrated a positive approach to students. | {1,2,3,4,5} |
| Q22 | The Instructor was open and respectful of the views of students about the course. | {1,2,3,4,5} |
| Q23 | The Instructor encouraged participation in the course. | {1,2,3,4,5} |
| Q24 | The Instructor gave relevant homework assignments/projects, and helped/guided students. | {1,2,3,4,5} |
| Q25 | The Instructor responded to questions about the course inside and outside of the course. | {1,2,3,4,5} |
| Q26 | The Instructor's evaluation system (midterm and final questions, projects, assignments, etc.) effectively measured the course objectives. | {1,2,3,4,5} |
| Q27 | The Instructor provided solutions to exams and discussed them with students. | {1,2,3,4,5} |
| Q28 | The Instructor treated all students in a right and objective manner. | {1,2,3,4,5} |

## IV. DATA PREPROCESSING

### A. Elimination of irrelevant attributes

Because the data analysis has not been done yet, it seems that all the attributes present in the data mining process is required.

### B. Grouping of attribute values

According to purpose of this paper, it was necessary nb.repeat field had to be chosen as the "class" attribute and the numerical values had to be replaced. Thus, if a student attends the relevant course more than one, Yes value will be replaced with that number (can be 2 or greater) otherwise, the field will be initialized with No. Because the attribute values of ID instructor (instr) and course code (class) are Nominal and may be due to numeric values incorrectly contribute as numeric attributes (Numerical) in the data mining algorithms, hence we replaced instr attribute values which are displayed with numbers 1, 2 and 3 with values A, B and C. The same operation for class attribute and for values of 1 to 13, with the alternative English letters from A to M is done. As stated earlier, the possible values for the attributes attendance, difficulty and Q1 to Q28 are 1,2,3,4 or 5 (except attendance that has the value 0). Grouping plan of these values are in such that values 1, 2 are replaced with Low, values 3 are replaced with Middle and values 4,5 are replaced with High (Zero is placed on the attendance attribute for the values 0).

## V. DATA ANALYSIS

To obtain a more accurate result, three types of evaluation has been performed on the data and subsequently, the data in accordance with the purpose of the special analysis is selected. Obviously, in all analyses, class field (nb.repeat) always be present. Analyses been carried out with the aid of data mining tool WEKA (Waikato Environment for Knowledge Analysis) [4].

### A. The first analysis

The goal is to investigate the role of providing a specific course by an instructor on not taking that course again by students.

To evaluate this analysis attributes class and instr have been selected.

By applying Apriori algorithm a number of valuable law is obtained:

```
1. class=E  ==> nb.repeat=No
2. instr=C  class=E  ==> nb.repeat=No
3. instr=A  ==> nb.repeat=No
```

Rule 1: Students has passed course E by taking it once.
Rule 2: Students has passed course C provided by instructor E by taking it once.
Rule 3: Students has passed the course that provided by instructor A by taking it once.

### B. The second analysis

The goal is to investigate the features of provided course on not taking that course again by students.

To evaluate this analysis attributes attendance, difficulty and Q1 to G12 have been selected. Among classification algorithms that is applied to the selected data, the best result provided by REPTree algorithm with the following features and decision tree:

| Correctly Classified Instances | 84.244 % |
|---|---|
| Avg. F-Measure | 0.774 |

```
REPTree
============

attendance = zero
|   difficulty = high
|   |   Q5 = middle
|   |   |   Q8 = middle : No
|   |   |   Q8 = high
|   |   |   |   Q3 = middle : No
|   |   |   |   Q3 = high : Yes
|   |   |   |   Q3 = low : Yes
|   |   |   Q8 = low
|   |   |   |   Q3 = middle : Yes
|   |   |   |   Q3 = high : No
|   |   |   |   Q3 = low : No
|   |   Q5 = high
|   |   |   Q9 = middle : No
|   |   |   Q9 = high
|   |   |   |   Q10 = middle : Yes
|   |   |   |   Q10 = high : No
|   |   |   |   Q10 = low : No
|   |   |   Q9 = low : No
|   |   Q5 = low : No
|   difficulty = middle : No
|   difficulty = low : No
attendance = low
|   Q2 = middle
|   |   Q1 = middle
|   |   |   Q9 = middle : No
|   |   |   Q9 = high
|   |   |   |   Q11 = middle : No
|   |   |   |   Q11 = high
|   |   |   |   |   Q4 = middle : No
|   |   |   |   |   Q4 = high : No
|   |   |   |   |   Q4 = low : Yes
|   |   |   |   Q11 = low : Yes
|   |   |   Q9 = low : No
|   |   Q1 = high : No
|   |   Q1 = low : No
|   Q2 = high : No
|   Q2 = low
|   |   Q7 = middle : No
|   |   Q7 = high : No
|   |   Q7 = low
|   |   |   difficulty = high : No
|   |   |   difficulty = middle
|   |   |   |   Q12 = middle : No
|   |   |   |   Q12 = high : Yes
|   |   |   |   Q12 = low
|   |   |   |   |   Q5 = middle : No
|   |   |   |   |   Q5 = high : No
|   |   |   |   |   Q5 = low
|   |   |   |   |   |   Q1 = middle : No
|   |   |   |   |   |   Q1 = high : Yes
|   |   |   |   |   |   Q1 = low : No
|   |   |   difficulty = low : No
attendance = middle : No
attendance = high : No
```

As it is deduced from the decision tree, the first and the most important attribute is attendance and highlights the fact that if a student be present continuously or middle in the classroom, he will pass the course with taking it once and otherwise attribute difficulty will be checked. If the lesson into student's opinion is simple or intermediate it will be passed with once taken. Considering the number of instances that meet the above conditions it seems these two attributes have the most important impact on not taken this course by the student more than once.

### C. The third analysis

The goal is to investigate the features of an instructor on not taking more than once the course provided by that instructor. To evaluate this analysis attributes Q1 to G12 have been selected. Among classification algorithms that is applied to the selected data, the best result provided by REPTree algorithm with the following features and decision tree:

| Correctly Classified Instances | 84.2612 % |
|---|---|
| Avg. F-Measure | 0.772 |

```
REPTree
============

Q14 = middle : No
Q14 = high : No
Q14 = low
|   Q16 = middle : No
|   Q16 = high
|   |   Q25 = middle : No
|   |   Q25 = high
|   |   |   Q15 = middle : No
|   |   |   Q15 = high : Yes
|   |   |   Q15 = low : No
|   |   Q25 = low : Yes
|   Q16 = low
|   |   Q25 = middle : No
|   |   Q25 = high : No
|   |   Q25 = low
|   |   |   Q26 = middle : No
|   |   |   Q26 = high : No
|   |   |   Q26 = low
|   |   |   |   Q13 = middle :
|   |   |   |   Q13 = high : No
|   |   |   |   Q13 = low
|   |   |   |   |   Q23 = middle : No
|   |   |   |   |   Q23 = high : No
|   |   |   |   |   Q23 = low
|   |   |   |   |   |   Q20 = middle
|   |   |   |   |   |   |   Q21 = middle : No
|   |   |   |   |   |   |   Q21 = high : No
|   |   |   |   |   |   |   Q21 = low : Yes
|   |   |   |   |   |   Q20 = high : No
|   |   |   |   |   |   Q20 = low : No
```

As it is deduced from the decision tree, the first and the most important attribute is Q14 and highlights the fact that if the instructor be fully or middle prepared to teach prior to entering the classroom the student will pass that course with taking it once and otherwise Q16 attribute will be checked. If the Instructor is committed to the course, the probability of student success will be very high and in the absence of instructor commitment to the course Q25 attribute will be checked and indicates that if the Instructor responded to questions about the course inside and outside of the course, chance of once getting a course by the student will be immense.

In case, the lack of instructor accountability outside of the course, Q26 attribute will be checked and indicates this point that if the Instructor's evaluation system effectively measured the course objectives, probability to once obtain that course by student is much more than a state where instructor's evaluation has any other purpose. Considering the number of instances that meet the above conditions it seems these four attributes have the most important impact on not been taken that course provided by this instructor more than once.

## VI. CONCLUSION

In this paper with using data mining techniques "Classification" (REPTree) and "Association rule discovery" (Apriori algorithm) effective factors on non re taking a course by students was investigated. 3 types of evaluations were performed on the data. The first analysis showed that if a particular course offered by a particular instructor students will pass that course by taking it once. In the second analysis the conclusion was that if a student continually be present in the classroom or course for students is easy to understand, that course will be passed on once taking and in the third analysis was indicated that if the instructor is fully or middle prepared to teach prior to entering the classroom and is committed to the course and if the Instructor responded to questions about the course inside and outside of the course and if the Instructor's evaluation system effectively measured the course objectives and not something else, then his students will pass the course provided by this instructor by once taking.

## VII. FUTURE WORK

As mentioned before in the data set used in this paper, performance of only three teachers has been studied and the results have been based on that belief, when these instances are not a suitable statistical community to express a general result and extend it to other states and instructors. However, it should also be noted that with this number of instructors a massive database has been made. If someone is willing to have broader and more credible research on affecting factors on academic performance of students based on the features of this dataset, it is recommended that he applies to collect and create a new data set personally although it is very time consuming and exhausting.